\documentclass[a4paper]{article}
\usepackage{amsfonts}
\usepackage{amsbsy}
\usepackage{amsmath}

\newcommand{\no}{\nonumber\\}
\newcommand{\p}{\mathbb P}
\renewcommand{\wp}{{\cal P}}

\begin{document}

\title{\bf{Multiple-event probability in \\ general-relativistic quantum mechanics}}

\author{\large
Frank Hellmann${}^{ab}$, Mauricio Mondragon${}^{b}$, Alejandro
Perez${}^{b}$, Carlo Rovelli${}^b$
 \\[3mm]
\em
\normalsize
{${}^a$Fakult\"at f\"ur Physik,  Ludwig-Maximilians-Universit\"at, D-80799 M\"unchen EU}
\\ \em\normalsize{${}^b$Centre de Physique Th\'eorique de Luminy%
\footnote{Unit\'e mixte de recherche (UMR 6207) du CNRS et des
Universit\'es de Provence (Aix-Marseille I), de la Mediterran\'ee
(Aix-Marseille II) et du Sud (Toulon-Var); laboratoire affili\'e \`a
la FRUMAM (FR 2291).}, Universit\'e de la M\'editerran\'ee, F-13288
Marseille EU}}

\date{\small\today}
\maketitle

\begin{abstract}

\noindent
We discuss the definition of quantum probability in the context of
``timeless" general--relativistic quantum mechanics. In particular,
we study the probability of
{\em sequences} of events, or multi-event probability. In
conventional quantum mechanics this can be obtained by means of the
``wave function collapse" algorithm.  We first point out certain difficulties
of some natural definitions of multi-event probability, including
the conditional probability widely considered in the literature.  We
then observe that multi-event probability can be reduced to
single-event probability, by taking into account the quantum nature
of the measuring apparatus.  In fact, by exploiting the von-Neumann
freedom of moving the quantum/classical boundary, one can always
trade a sequence of \emph{non-commuting} quantum measurements
at different  times, with an ensemble of simultaneous \emph{commuting}
measurements on the  joint system+apparatus system.
This observation permits a formulation
of quantum theory based only on single-event probability, where
the results of the  ``wave function collapse" algorithm can
nevertheless be recovered. The discussion bears also
on the nature of the quantum collapse.
\end{abstract}

\section{Introduction}

The general-relativistic revolution of our understanding of space
and time has proven extremely effective empirically.
Conventional textbook quantum mechanics (QM), and
conventional quantum field theory (QFT), however, are formulated in a language
which is incompatible with the general-relativistic notions of
space and, especially, time.  Is there a
formulation of QM compatible with these notions? Such a
formulation should be required, in particular, in order to provide a clear interpretative
framework  to any attempt to formulate a quantum theory of gravity
in a form consistent with the general-relativistic understanding of
space and time \cite{Kiefer}.

The main difficulty in extending QM to the general-relativistic
context comes from the notion of time. (Historical, if a bit
out-of-date reviews are \cite{Kuchar:1991qf,Isham:1992ms}. See also
the recent \cite{Dittrich:2004cb}  and \cite{Dolby:2004ak}.) In
conventional QM (including QFT), the independent time variable $t$
is interpreted as an observable parameter independent from the
physical system considered. States and probability amplitudes evolve
deterministically in $t$.\footnote{In QFT this is realized via the
representation of the Poincar\'e group in the state space.} This
notion of time is not general-relativistic and conflicts with
general covariance, and so, if we take general relativity seriously, it is
likely to be unsuitable to describe the world at the Planck scale.

In a sum-over-histories context, this problem is addressed
by Hartle's generalized QM \cite{Hartle:1992as}.
Here we work in the hamiltonian context.  We summarize below,
in Section \ref{QM}, a hamiltonian formulation of QM that does not
make use of this nonrelativistic notion of time, following \cite{covariant}, and
especially Chapter 5 of \cite{book}, to which we refer for  full details and a complete
discussion. Following  \cite{book}, we call this formulation ``general-relativistic".
In this formulation, the probability
for observing a certain event  $s'$ if an event $s$
was observed, is given by the modulus square of a suitably defined
transition amplitude  $\wp_{s \Rightarrow s'}=|{\cal A}_{s{\Rightarrow} s'}|^2
=|\langle s' | s \rangle_{\cal H}|^2$, as explained below. This probability postulate coincides
with standard QM transition probabilities in the
non-general-relativistic case, but is well-defined in a
wider  context, sufficiently general to form the basis for a timeless
interpretation of a general--relativistic quantum theory.

However, a key problem was left open in \cite{book}.  QM does not
provide only probabilities for single observations.  It also
provides probabilities for sequences of observations. For instance,
it provides the probability $\wp_{\psi{\Rightarrow} \psi'\psi''}$
for a system in a state $|\psi\rangle$ to be observed in a state
$|\psi'\rangle$ and then in a state $|\psi''\rangle$. One way to
compute this is to assume that at the time of the first measurement
the state ``collapses" to the state $|\psi'\rangle$. It is not clear
how these probabilities for ensembles of events can be computed in
general-relativistic QM. The difficulty comes from the following
fact.  As is well known, in QM, given a un-ordered ensemble of
observations, its probability depends on the \emph{time order} in
which they are performed. Now, in a quantum general relativistic
context, as we recall below, the notion of time-evolution, and in
particular the notion of time-order, is subtle due to the absence of
an external notion of time. How do we compute, then, the probability
of sequences of measurements?  The problem was discussed for
instance by Hartle in \cite{Hartle}.  A strictly related problem is
that the probability defined in \cite{book} concerns only the
\emph{non-degenerate} eigenvalues of the (partial) observables. The
probability of observing a \emph{degenerate} eigenvalue was not
defined.  In this paper we study a possible solution to this
problem.

In Section \ref{3}, we state the problem and we
point out a certain number of difficulties
that emerge in trying to assign a probability to sets of events
in a general relativistic context.  In particular, we discuss the
difficulties of two apparently ``natural" solutions. The first is a
direct generalization of the single event probability postulate: the
probability of an ensemble of events is determined by
the projection on the physical Hilbert space of the subspace of the
kinematical Hilbert space associated to this ensemble of events. We
show that this postulate is not viable because it does not reduce to
the standard QM probabilities in the nonrelativistic case.   The second
is the use of conditional probability, widely discussed in the literature.
We point out certain difficulties with the operational definition of this probability.

In Section \ref{apparatus}, we indicate a possible general way for
solving the problem.   This is based on the observation that a
multiple-event probability, such as $\wp_{\psi{\Rightarrow}
\psi'\psi''}$ can always be reinterpreted as a single-event
probability, once the dynamics and the quantum nature of the
apparatus making the measurements are taken into account.  If we do
so, the time order gets naturally coded into the dynamics of the
system.  This strategy provides a general way for dealing with
multiple-event probabilities in general-relativistic quantum
mechanics.

Finally, in Section \ref{probabilita}
we comment on the meaning of the notion of probability in
the timeless case. In particular, we clarify the apparent
difficulty presented by  the fact that probabilities assigned to the possible values
of a variable may not sum up to one.
In Section \ref{fine} we summarize our results and discuss the issues that
remain open.

We illustrate the working of this technique in a companion
paper \cite{model}, where we introduce a discrete model that allows
us to exemplifies all the general structures introduced above.
The example in  \cite{model} illustrates a notable convergence
between the strategy introduced here and Hartle's generalized QM
approach  \cite{Hartle:1992as}.

\section{General relativistic quantum mechanics}\label{QM}

We give a rapid overview of general relativistic QM,
referring to Chapter 5 of  \cite{book} for a complete discussion and
full details.

\subsection{General relativistic classical mechanics}\label{CM}

A classical mechanical system is described by
a number of observables quantities, that we call \emph{partial
observables}.  We include under this name the dependent as well as
the independent variables in the equations of motion of
non-relativistic mechanics.  Examples of partial observables are the
time $t$, the position of a particle $\vec X$, the momentum, the energy,
the spacetime coordinates $x^\mu$ in special--relativistic field theory, the four
coordinates $x^\mu$ of a relativistic particle, and so on. We assume
that each partial observable can be measured by  a suitable physical
apparatus.

In a system with a finite number $n$ of degrees of freedom, we
choose $n+1$ partial observables (typically the $n$ lagrangian
variables plus the time variable), and form the $n+1$ dimensional
\emph{extended configuration space}, or event space, $\cal C$.   The
extended configuration space of a relativistic particle is the
Minkowski space.   The extended configuration space of a homogenous
and isotropic cosmological model where $a$ is the volume of the
universe and $\phi$ is the matter density, is coordinatized by $a$
and $\phi$. Points in $\cal C$ are called \emph{events} and denoted
$s,s',s'',...$: for instance, a point in Minkowski space
$s=x^\mu=(\vec x, t)$, or a given value of radius of the universe
and matter density $s=(a, \phi)$, define an event.   Measuring a
complete set of partial observables --that is, determining a point
in $\cal C$-- is to detect the happening of a certain event. For
instance: a particle is detected in a point of Minkowski space, a
certain value of radius of the universe and average energy density
are measured, and so on. Each such detection describes an
interaction of the system with another system, playing the role of
observer.

Dynamics is then uniquely determined by a single function $H$, the
relativistic hamiltonian\footnote{It is also called ``super--hamiltonian",
``hamiltonian constraint", or ``scalar constraint".}, on the
cotangent space $T_*{\cal C}$.\footnote{In this paper, we do not
consider the case of other gauge invariances, besides the reparametrization
invariance generated by $H$.}  This is a symplectic space, with a
natural symplectic two-form $\omega$.    The submanifold  $\Sigma$ defined by
$H=0$ is called the constraint surface.  The integral lines
of the restrictions of $\omega$ to $\Sigma$ are the physical motions
of the system.   The space of these motions
is the phase space $\Gamma$.
Thus, in this context the phase space $\Gamma$ is interpreted, \`a
la Lagrange, as the space of the solutions of the equations of
motion\footnote{More precisely, as the space of the equivalence
classes of solutions, up to gauge.},
or the space of the possible ``motions" of the system, instead than
a space of initial data. Notice that a physical motion determines
a continuous sequence of events in $\cal C$.

The interest of this formulation of dynamics is that the time
variable $t$ is just one of the coordinates of $\cal C$, on equal
footing with the other partial observables, as required by general
covariance.  In a general covariant system, in fact, dynamics is not
the description of how physical variables evolve in a preferred
independent variable $t$, but rather the description of the
physical relations between partial observables (see the discussion in
\cite{book}, Sections 3.1 and 3.2.4).

A special case is provided by the non-relativistic systems.  In
these systems, one of the coordinates of $\cal C$ is singled out to
play a special role: the time variable $t$, thus ${\cal C}={\cal
C}_0\times \mathbb R$, where ${\cal C}_0$ is the conventional
configuration space; and $H$ has the form
 \begin{equation}
 H=p_t + H_0
\label{nonrelativistic}
\end{equation}
 where $p_t$ is the
momentum conjugate to $t$ and $H_0$ is independent from $p_t$. $H_0$
is the standard non-relativistic hamiltonian. It is easy to see that
in this case $\Sigma$ is isomorphic to $\mathbb R\times T_*{\cal
C}_0$ and the motions are the integral lines of the hamiltonian flow
of $H_0$ on $T_*{\cal C}_0$: that is, we recover conventional
hamiltonian mechanics. Our main interest, of course, is for systems
where $H$ does \emph{not} have the form (\ref{nonrelativistic}).
These are the systems for which conventional QM is insufficient, and
general relativity is among (the $\infty$-number of degrees of
freedom version of) these systems.

\subsection{General relativistic quantum mechanics: basic structure}\label{CM1}

General relativistic quantum mechanics is defined by a kinematical
Hilbert space $\cal K$ carrying an algebra of operators
corresponding to the partial observables.  The dynamics is given by
a generalized projection operator $\p$ in $\cal K$.

The relation with a classical system is as follows.   A linear
representation of the Poisson algebra of coordinates and momenta on
$T_*{\cal C}$ defines the kinematical Hilbert space $\cal K$. For
instance, we may take a Schr\"odinger representation ${\cal
K}=L_2[{\cal C}]$.  The relativistic hamiltonian $H$ defines a
self-adjoint quantum operator (that we denote with the same symbol)
$H$ on $\cal K$. The kernel of $H$, formed by the (possibly
generalized) states in $\cal K$ satisfying the ``Wheeler-DeWitt"
equation
\begin{equation}
 H\psi=0 \label{wdw} \end{equation} is called the physical state space
$\cal H$.  $\p$ is the linear (self-adjoint) operator $\p: {\cal
K}{\to}{\cal H}$, given by:
\begin{equation}
\label{projector}
\p\psi=\int dn\ e^{-inH}\ \psi
\end{equation}
(we put $\hbar=1$) and loosely called ``the projector", since
$\psi\perp{\cal H}\Longleftrightarrow \p\psi=0$.\footnote{If ${\cal
K}/{\cal H}$ has finite dimension $N$ then $\p=N\tilde \p$, with
$\tilde \p$ the true projector from $\cal K$ to $\cal H$.} For
periodic systems the range of integration is the period, for the
others it is the real line (recall that we are only working with
reparametrization invariant systems). If zero is in the discrete
(resp.\ continuous) spectrum of $H$, then $\cal H$ is a proper
(resp.\ generalized) eigenspace of $\cal K$. On the linear space
$\cal H$ we consider the Hilbert structure
\begin{equation}
\langle \p s' |\p s \rangle_{\cal H} := \langle s' |\p| s\rangle \ ,
\end{equation}
which is well defined (on a dense subspace of $\cal K$) even when
$\cal H$ is a generalized eigenspace; see for instance
\cite{Marolf:2000iq} and
 the discussion in
Section 5.5.2 of \cite{book}. We also write $\langle s' |s \rangle_{\cal H}\equiv
\langle \p s' |\p s \rangle_{\cal H}$.  Remark that since in general $\p$ is not
a true projector, $\langle s' |\p| s\rangle$ may very well be different from
$\langle\p s' |\p s\rangle$. In particular, this last quantity is in general
divergent in the case in which $\cal H$ is a generalized eigenspace.

States in $\cal K$ have a physical interpretation\footnote{This is
the distinguishing feature of the interpretation we are
considering.}, as follows. If $|s\rangle\in\cal K$ is an eigenstate
of a complete set of commuting partial observable (self-adjoint)
operators, with eigenvalues $(a,b,c...)$, then $|s\rangle$ is
interpreted as describing the event in $\cal C$ with coordinates
$(a,b,c...)$. That is, it describes an interaction of the system
with another physical system, in which the values $(a,b,c...)$ are
realized.

In a non-relativistic system, where (\ref{nonrelativistic}) holds,
the Wheeler--DeWitt equation (\ref{wdw}) becomes the Schr\"odinger
equation and $\p$ is strictly related to the unitary evolution
operator $U(t)=e^{-iH_0t}$.      As a simple example, consider a
2-state spin system with time--independent hamiltonian $H_0$.  Here
${\cal K}= \mathbb C^2\otimes L[\mathbb R]$ is formed by states of
the form $\psi_S(t), S=\uparrow, \downarrow$ and spanned by the
(generalized) basis $|S, t \rangle$, where $\psi_S(t)= \langle S,
t|\psi\rangle $, and
\begin{equation}
\p| S, t \rangle=\int_{-\infty}^\infty
d\tau\ e^{-i\tau(p_t+H_0)}\ | S, t \rangle =  \int_{-\infty}^\infty
d\tau\ e^{-iH_0\tau}\ | S, t+\tau \rangle.
\label{def}
\end{equation}
$\p$ sends $| S, t \rangle$ into a solutions of the Schr\"odinger
equation. The physical space $\cal H$ is given by these solutions
and clearly ${\cal H}\sim \mathbb{C}^ 2$ (any initial state
generates one and only one solution of the Schr\"odinger equation).
A basis in $\cal H$ can be obtained by choosing a reference time
$t_0$ and defining $|S\rangle:= \p| S, t_0\rangle$. It follows
easily $\p|S, t \rangle= U^\dagger(t-t_0)|S\rangle$. Then notice
that
\begin{equation}
\langle S',t' |\p |S, t \rangle=  \langle S' |e^{-iH_0(t'-t)}|S\rangle,
\label{spinproj}
\end{equation}
that is, the matrix elements of $\p$ are essentially the matrix elements of
the unitary evolution operator.

\subsection{Single event probability}\label{single}

The main interpretation postulate of general relativistic QM is the
following (see \cite{book}). The probability $\wp_{s{\Rightarrow}
s'} $ of observing the event $s'$ if the event $s$ was observed (we
shall write $\wp_{s{\Rightarrow} s'} $ simply as $\wp_{s'}$ when
there is no need of indicating the initial state) is given by  the
modulus square of the amplitude
\begin{equation}
            A_{s{\Rightarrow} s'} = \langle s' |\p| s \rangle,
  \label{singleevent}
  \end{equation}
where the states are normalized in $\cal H$, not in $\cal K$, that is
\begin{equation}
              \langle s |\p| s\rangle = \langle s' |\p| s' \rangle= 1.
              \label{normalization}
\end{equation}

There are several equivalent ways of writing this probability.
We can define the projection operator
\begin{equation}
              \Pi_{s'} \ \  = \ \  \p  | s' \rangle\, \langle  s' | \p \ \  = \ \   | \p s' \rangle\, \langle \p s' |
            \label{singleeventP}
\end{equation}
and write the probability as the expectation value
\begin{equation}
            \wp_{s{\Rightarrow} s'} =  |A_{s{\Rightarrow} s'}|^2 =  \langle s |     \Pi_{s'}   |s \rangle,
            \label{singleeventnorm}
\end{equation}
where again  $|s \rangle$ and  $|s' \rangle$ are normalized by (\ref{normalization}).
Alternatively, recalling the notation
 $\langle s' |s \rangle_{\cal H}\equiv \langle \p s' | \p s \rangle_{\cal H}:=
\langle s' |\p| s\rangle$, and explicitly including the normalization in the expression for the
probability, we can write
\begin{equation}
            \wp_{s{\Rightarrow} s'} \  =\
            \frac{| \langle s' |s \rangle_{\cal H}|^2}{ \langle s' |s' \rangle_{\cal H} \
            \langle s |s \rangle_{\cal H}}
           \   =\
            \frac{| \langle \p s' |\p s \rangle_{\cal H}|^2}{ \langle \p s' | \p s' \rangle_{\cal H} \
            \langle \p s | \p  s \rangle_{\cal H}} .
            \label{singleevent3}
\end{equation}

Notice that this probability is a standard quantum mechanical
probability computed in the \emph{physical} Hilbert space $\cal H$,
in the following sense. The states $|s \rangle$ and $|s' \rangle$ in
$\cal K$ ``project'' down to physical states $|\p s \rangle$ and
$|\p s' \rangle$ in $\cal H$. The probability (\ref{singleevent3})
is then simply the standard probability amplitude of measuring the
physical state $|\p s' \rangle$ if the physical state $|\p s\rangle$
was measured. In other words, $  \Pi_{s'}$ is the projector onto
$|\p s' \rangle$, in $\cal H$, and it is a \emph{genuine Dirac
operator}.  Therefore general relativistic QM simplifies, but does
not contradict, the standard Dirac treatment of a constrained
system. This fact, by the way, assures us that the amplitude
(\ref{singleevent}) yields a probability less or equal to one.

This formulation is very close in spirit
to the conventional scattering formulation of QFT, where probabilities are
defined by transition amplitudes between initial and final
asymptotic states, and these are defined as timeless Heisenberg
states, or full ``histories" of the system: see for instance
Weinberg's  clear discussion in
Chapter 3 of \cite{weinberg}.

This definition of probability reduces to the conventional one in
the non-relativistic case. In the spin system considered above, for
instance, the states $\p |S,t\rangle$ are normalized and the
amplitude for measuring the spin $S'$ at time $t'$ if $S$ at time
$t$ was measured is
\begin{equation}
            A_{St\; \Rightarrow\; S't'} = \langle S',t' |\p| S,t \rangle
            = \langle S' |e^{-iH_0(t'-t)}|S\rangle,
\end{equation}
in agreemenet with conventional QM.
More in general, in the case in which $H=  p_t+H_0$, the definitions
above reduce to the standard QM postulates regarding states,
observables and probability.

However, there is one additional postulate required to define quantum
theory: the collapse postulate, stating that \emph{after} a
measurement, the state changes and becomes an eigenstate of the
operator associated to the measurement. The translation of this
postulate to the relativistic formalism is our concern here.

\section{Multiple-event probability}\label{3}

Consider a partial observable $  A$ in $\cal K$ and let $a$ be
one of its eigenvalues. If $a$ is non-degenerate, and $|s'\rangle$
is the corresponding eigenstate, then (\ref{singleevent}) provides
the probability amplitude of measuring $a$.  What happens if $a$ is
degenerate?

Let us say for simplicity that $a$ is doubly degenerate, and that
$|s'\rangle$ and $|s''\rangle$ are two orthogonal eigenstates having
eigenvalue $a$, that is, they span the $a$-eigenspace ${\cal K}_a$.
Then, to measure the eigenvalue $a$, or, equivalently, to measure
its associated projection operator $\pi_a =|s'\rangle\langle
s'|+|s''\rangle\langle s''|$, means that we have a measuring
apparatus that gives us a Yes answer if either the event $s'$ or the
event $s''$ happen (Yes answer corresponds to the eigenvalue 1 of
$\pi_a$). In order to compute the probability of having a Yes
answer, we need therefore the probability $\wp_{s'\,
{\scriptscriptstyle \rm OR}\, s''}$ that the event $s'$ OR the event
$s''$ happens.

Alternatively, suppose that we have a measuring apparatus that gives
us a Yes answer if both the event $s'$ and the event $s''$ happen.
In order to compute the probability of having a Yes answer, we need
therefore the probability $\wp_{s'\, {\scriptscriptstyle \rm AND}\,
s''}$ that the event $s'$ AND the event $s''$ happen.    The
solution of one case gives immediately the solution of the other
since, clearly
\begin{equation}
\wp_{s'\, {\scriptscriptstyle \rm OR}\, s''} =
\wp_{s'} + \wp_{s''} - \wp_{s'\, {\scriptscriptstyle \rm AND}\,
s''}.
\label{orand}
\end{equation}
There are two possibilities: either
$\wp_{s'\, {\scriptscriptstyle \rm AND}\, s''}$ is always zero, or
not.  Consider the two cases separately.

\paragraph{\em (i) Mutually exclusive events.} If $\wp_{s{\Rightarrow} (s'\, {\scriptscriptstyle \rm AND}\, s'')}=0$
for any $s$, then $s'$ and $s''$ are alternative events that cannot
both happen.  That is, if one happens, the probability that the
other happens is zero.  By
 (\ref{singleevent}) and \emph{the given interpretation}, this is
equivalent to
\begin{equation}
        \langle s' |    \p   |s'' \rangle = 0.
\label{orto1}
 \end{equation}
In this case, (\ref{orand}) gives
\begin{equation}
\wp_{s'\,{\scriptscriptstyle \rm OR}\, s''} = \wp_{s'} +  \wp_{s''}.
\label{orto}
 \end{equation}
That is, the probability of $s'$ OR $s''$ is simply the sum of the
probabilities of $s'$ and $s''$.  Observe that this can be written
generalizing  (\ref{singleeventnorm}) to
 \begin{equation}
            \wp_{s{\Rightarrow} a} =   \langle s |    \Pi_{a}   |s \rangle,
                                  \label{doubleeventP}
\end{equation}
where $ \Pi_{a}$ is the orthogonal projector on the subspace ${\cal
H}_a=\p{\cal K}_a$ in $\cal H$, which, if (\ref{orto1}) holds, is
simply given by (remember we are requiring \eqref{normalization})
 \begin{equation}
\Pi_a = |\p s'\rangle\langle \p  s'|+| \p s''\rangle\langle \p  s''|.
\label{troppobello}
  \end{equation}

A typical example is the following.  In the two-state spin system
considered in the previous section,
let $|s'\rangle=|\!\uparrow, t\rangle$ and $|s''\rangle=
|\!\downarrow, t\rangle$.   In this case, $\langle
s' |\p|s'' \rangle=\langle  \uparrow  |U(0)|\!\downarrow \rangle= 0$.
The two events are mutually exclusive.  Therefore
$\wp_{s'\, {\scriptscriptstyle \rm AND}\, s''}=0$. The projector
on the $a$-eigenspace ${\cal K}_a$ is
\begin{equation}
 \pi_{a}=   |\!\uparrow\, , t \rangle\langle\,\uparrow, t |\ +\
 |\!\downarrow\,, t\rangle\langle\,\downarrow , t|\ .
\end{equation}
The projection ${\cal H}_a$ of ${\cal K}_a$ in ${\cal H}$ is spanned
by the two orthogonal states $\p\,|\!\!\uparrow, t\rangle
=U^\dagger(t-t_0)|\!\uparrow\,\rangle$ and $\p\,|\!\downarrow,t \rangle
=U^\dagger(t-t_0)|\!\downarrow\,\rangle$, therefore
\begin{equation}
 \Pi_{a}\ =\ U^\dagger(t-t_0)\ \Big(  |\!\uparrow\, \rangle\langle\,\uparrow\!|\ +\
 |\!\downarrow\,\rangle\langle\,\downarrow\!|\  \Big)\ U(t-t_0).
\end{equation}
In this two-state system, $\Pi_{a}=1\!\!1$ and the corresponding
probability is $\wp_a=1$. Not so, of course, in general.

\paragraph{\em (ii) Non exclusive events.}  The interesting case is when
\begin{equation}
              \langle s'   |\p|s'' \rangle \ne 0.
                                  \label{nonidependent}
\end{equation}
A typical example is the following.  Let $|s'\rangle=|\!\uparrow,
t'\rangle$ and $|s''\rangle= |\!\leftarrow,
t''\rangle=\frac{|\uparrow, t''\rangle+|\downarrow,
t''\rangle}{\sqrt2}$.   In this case, $\langle s'' |\p|s'
\rangle=\langle\, \leftarrow  |U(t''-t')| \!\uparrow \rangle \ne 0$,
in general. The question we are asking is: what is the probability
of detecting the spin  $\uparrow$ at time $t'$ AND the spin
$\leftarrow$ at $t''$? The problem is of course well posed: if a
particle is in a certain initial state at $t$, what is the
probability of finding it with a certain spin at time $t'$ AND with
another spin at a later time $t''$? This can be measured by
measuring the fraction of a beam that passes through a sequences of
two Stern-Gerlach apparatuses.

Now, in ordinary quantum mechanics, these probabilities depends on the
\emph{time ordering} between the events.   For instance, let the
initial state $|s\rangle$ be the state
$|{\rightarrow}\rangle=\frac{|\uparrow\rangle-|\downarrow
\rangle}{\sqrt2}$ at time $t_0$, and let $U(t)=1\!\!1$ for all $t$. Then
\begin{equation}
\wp_{s{\Rightarrow}(s'\, {\scriptscriptstyle
\rm AND}\, s'')}\ =
\begin{cases}\frac{1}{4} \hspace{1em}& {\rm if}\  \  t'<t'' \     , \\
 0 \hspace{1.4em} & {\rm if}\ \  t''<t'\  .
\end{cases}
\label{jenny}
\end{equation}
Because the sequence
\begin{equation}
 |\! {\rightarrow}\rangle \ \  \ \ \Rightarrow \ \  \ \   |\!\uparrow\rangle  \ \ \ \
 \Rightarrow \ \ \ \
|\!\leftarrow\rangle
\label{sequence}
\end{equation}
has probability
1/4; while the sequence
\begin{equation}
 |\! {\rightarrow}\rangle\ \    \ \  \Rightarrow \ \  \ \   |\!\leftarrow\rangle  \ \  \ \  \Rightarrow  \ \  \ \
|\!\uparrow\rangle
\label{sequence2}
\end{equation}
cannot happen.
The standard way of obtaining these probabilities in conventional
quantum mechanics is via the projection postulate.  For instance,
say $t'<t''$, that is, case (\ref{sequence}). We have: $(i)$ at time
$t'$ the spin $\uparrow$ is measured with probability $| \langle
\uparrow | \leftarrow\rangle|^2=1/2$; $(ii)$ the state is hence
projected to $ |\!\!\uparrow \rangle$; $(iii)$ at time $t''$ the spin
$\leftarrow$ is measured with probability $| \langle \leftarrow |
\uparrow\rangle|^2=1/2$, giving total probability $1/2\times
1/2=1/4$.

Standard QM gives also, easily
\begin{equation}
\label{22extra}
\mathcal{P}_{s\Rightarrow (s'\, {\scriptscriptstyle
\rm OR}\, s'')}
=\begin{cases}
  \frac{3}{4}  \hspace{1em} & \textrm{if $t'<t''$},  \\
  \frac{1}{2} & \textrm{if $t''<t'$}.  \\
  \end{cases}
 \end{equation}
Comparing with (\ref{orand}), notice that the probabilities  $\mathcal{P}_{s\Rightarrow s'}$
and  $\mathcal{P}_{s\Rightarrow s''}$ relevant here (with two detectors) are different from the probabilities
$\mathcal{P}_{s\Rightarrow s'}$ and  $\mathcal{P}_{s\Rightarrow s''}$ relevant when
only one detector is present.  For instance, in the first case, we have
$\mathcal{P}_{s\Rightarrow s''}=|\langle \rightarrow|\uparrow\rangle\langle\uparrow| \leftarrow\rangle|^2+|\langle \rightarrow|\downarrow\rangle\langle\downarrow|\leftarrow\rangle|^2=\frac12$, because of the presence of a detector in $s'$;
while in the absence of this, we would clearly have $\mathcal{P}_{s\Rightarrow s''}=|\langle \rightarrow|\leftarrow\rangle|^2 =0$.  This well known fact will play an important role below.

How do we recover these probabilities in relativistic QM, where we
do not have a notion of time ordering in $t$?

\subsection{Two false tracks}\label{false}

\subsubsection{Taking  (\ref{doubleeventP}) as the general probability postulate}

Let us suppose we ignore for the moment the time-ordering problem,
and we try to directly associate a probability to the $a$-eigenspace
${\cal K}_a$, as we did in the case {\bf\em (i)} (mutually exclusive events).
 Notice that the definition (\ref{doubleeventP}) of
probability remains meaningful also in the case
 {\bf\em (ii)} (non mutually exclusive events). It is therefore very tempting to suppose that
the probability is still given by (\ref{doubleeventP}) also in this
case.

In the example above, for instance, the eigenspace ${\cal K}_a$ is
spanned by the two events considered (which are orthogonal in
$\mathcal{K}$ because they are at different time), and the
projection operator on ${\cal K}_a$ is
\begin{equation}
 \pi_{a}= |\!\uparrow,t' \rangle\langle\!\uparrow,t'|+|\!\leftarrow,t'' \rangle\langle\!\leftarrow,t''|.
\end{equation}
The projection ${\cal H}_a$ of ${\cal K}_a$ to ${\cal H}$ is spanned
by the two states $\p|\uparrow, t'\rangle$
and $\p|\leftarrow,t'' \rangle$. This time these two states are not orthogonal in
$\cal H$. We can nevertheless still consider the possibility that the
probability $\wp_a$ is given by  (\ref{doubleeventP}), where
$\Pi_a$ is the projection operator on the space ${\cal H}_a$ they span.

The probability postulate  (\ref{doubleeventP}) is not of
completely straightforward utilization in the case in which the
orthogonal eigenstates of $\pi_a$ are projected to non-orthogonal
states in $\cal H$, namely in the case (\ref{nonidependent}),
because in this case (\ref{troppobello}) is not true.
This is a technical difficulty that can be
addressed by standard linear algebra methods, for instance via an
orthogonalization procedure. A more powerful technique is to
observe that the projector on the linear space spanned by a set of
possibly linearly dependent states $S=\{|u^1\rangle...|u^N\rangle\}$ can
be written in the form
\begin{equation}
 \Pi_{a}= \sum_{ij}\  |u^i\rangle\ G_{ij}\ \langle u^j |
\end{equation}
where the sum is over any maximal subset of linear independent sates
in $S$ and the matrix $G_{ij}$ is the inverse of their Gramm matrix
$G^{ij}=\langle u^i|u^j\rangle$.

Unfortunately, however, there is a serious problem: this
probability postulate is wrong because it does not reduce to the
conventional and well tested probabilities of nonrelativistic QM.
This can be seen from the following example. Consider the spin
case mentioned above.  The two kinematical states $|s'\rangle$ and
$|s''\rangle$  are orthogonal in $\cal K$ (because they are at
different times). If we project them down to $\cal H$, the two
resulting states $|\p s'\rangle$ and $|\p s''\rangle$ are not
anymore orthogonal, but they are still linearly independent.
Therefore they still span a two-dimensional space. Since the space
they span is ${\cal H}_a$, it follows that ${\cal H}_a$ is
two-dimensional.   But so is $\cal H$ in this example.  Hence
${\cal H}_a={\cal H}$. It follows immediately that $\Pi_{a} $ is
the identity operator in $\cal H$, and therefore that
(\ref{doubleeventP}) states that the probability of measuring
either $s'$ or $s''$ is always equal to unity. This is in
contradiction with the correct result (\ref{22extra})  given by
nonrelativistic QM.

Therefore the probability formula (\ref{doubleeventP}) is not
correct in the case {\bf\em (ii)} of compatible events.   The difficulty
appears to be in the fact that the formalism ignores that $s'$
happens \emph{before} $s''$.

\subsubsection{Conditional probabilities}

The idea that in the timeless case the interpretation of QM can be
entirely based on conditional probabilities has been suggested
in \cite{pw} and is very attractive.   The idea has been widely discussed and
also criticized, see
for instance \cite{Kuchar:1991qf}, but has recently received new
attention.  For instance, in \cite{Dolby:2004ak}, C. Dolby has provided
an interesting and convincing reply to the criticisms in
\cite{Kuchar:1991qf}.  We refer therefore here to Dolby's version of
the conditional probability interpretation. For full references, see the two
papers quoted.

Dolby gives a postulate for the probability $\wp(a\
{\rm when} \ b)$ that an event $a$ happens together with (as the
same time as) an event $b$.\footnote{Carefully distinguishing
$\wp(a\ {\rm when} \ b)$ from the conditional probability
$\wp(a\ {\rm if} \ b)$.}  The events $a$ and $b$ are described
by commuting projector operators $\pi_a$ and $\pi_b$ in $\cal K$.
Dolby's probability postulate is, in our notation (and not writing explicitly the
dependence of the probability on the state),
 \begin{equation}
 \wp(a\
{\rm when} \ b) =\frac{\langle s | \p \pi_a\pi_b \p  | s
\rangle}{\langle s |\p \pi_b \p  | s \rangle}.
\label{dolby}
\end{equation}
Suppose that we have a set $\{a_1, ..., a_n \}$ of events such
that
\begin{equation}
\sum_n \ \pi_{a_n}=1\!\!1.
\end{equation}
It follows immediately
from Dolby's definition that
\begin{equation}
\sum_n \ \wp(a_n\  {\rm when} \
b) = 1.
\label{dolbysum}
\end{equation}

In order to illustrate the difficulty with this definition of
probability,  consider the two-state system introduced in Section
\ref{CM1}, but let us imagine, for simplicity, that time is
discrete. That is, the states are  $\psi_S(t_n), S=\uparrow,
\downarrow$ where $\psi_S(t)= \langle S, t_n|\psi\rangle $, with
integer $n$ (as in the example we will discuss in the companion
paper \cite{model}).  In the conventional formalism one focus on
probabilities of the form $\wp(\uparrow {\rm when}\ t_n)$, where the
event $(\uparrow, t_n)$ is considered as one element of the set of
equal--time alternatives ${\cal S}_{t_n}=\{(\uparrow,
t_n)(\downarrow, t_n)\}$. But the general formalism does not
privilege the time variable and therefore allows us to consider also
probabilities of the form $\wp(t_n\ {\rm when} \uparrow)$, where the
event $(\uparrow, t_n)$ is considered as one element of the set of
alternatives ${\cal S}_\uparrow = \{..., (\uparrow,
t_{n-1}),(\uparrow, t_n),(\uparrow, t_{n+1}), ...\}$. Let us
therefore study the interpretation of these. If we take
$\pi_b=\pi_\uparrow$, the projector on the $\uparrow$ eigenspace of
the spin operator, and $\pi_{a_n}=\pi_{t_n}$, the projector on the
$t_n$ eigenspace of the $t$ operator in (\ref{dolby}), we have the
probability
 $\wp(t_n\ {\rm when} \uparrow)$  to find the particle at time $t_n$ ``when" the spin is $
\uparrow$. Let us calculate this probability for a state such that, in particular (assume
the dynamics is such that this is a physical state)
\begin{align}\label{condprob ex}
\psi_\uparrow(t_n)=\begin{cases} 1, & \mathrm{if} \quad n=1,2. \\
0, & \mathrm{otherwise.}
\end{cases}
\end{align}
There are two different times at which the spin is
$\uparrow$: $t_1$ and $t_2$. By symmetry and (\ref{dolbysum}), we
have immediately
\begin{equation}\label{condprob ex2}
\wp(t_1\  {\rm when} \uparrow)=\wp(t_2\  {\rm when}
\uparrow)=\frac12.
\end{equation}
At first sight, this looks
reasonable: since the particle has spin up at two different times,
if the spin is up, then there might be a 50/50 probability that it
is one or the other of these two times.  But can we give a more
precise definition of this probability? That is, can we give a
precise operational procedure for measuring this probability?
Here is where we see a difficulty. Indeed, we see two
possibilities for an operational interpretation of this
probability, but neither appears to work in general. The following are the two possibilities.

(i)  There is a single detector at $t=t_n$, that detects the spin $\uparrow$, and we interpret
$\wp(t_1\  {\rm when} \uparrow)$ as (the limit of) the ratio of the number of detections over the number of trials.   This is clearly not viable, because under the conditions given, this ratio is equal to one, not to 1/2: a detector which is on at time $t_1$ will {\em always} detect the particle, not just half of the times. How is this probability equal to one accounted for in Dolby's scheme?

(ii) We have one detector at each $t_n$, each sensitive only to the spin $\uparrow$.  We call $N_{t_n\uparrow}$ the number of times the detector at $t_n$ clicks, and call $N_\uparrow$ the total number of detections, and we interpret $\wp(t_1\  {\rm when} \uparrow)$ as the limit of $N_{t_1\uparrow}/N_\uparrow$. This works, because at each trial both detectors at $t_1$ and $t_2$ click, so that we get the correct 1/2.

However, (ii) does not work in general, because of the fact that, as noticed after
(\ref{22extra}), the very presence of detectors alters quantum mechanical probabilities.
For instance, suppose we have a system
with four states $ |i\rangle, i=1,2,3,4$, with a time dependent dynamics given by
$\psi(t_{n+1})={\mathcal U}(t_n)\psi(t_n)$ where
\begin{equation}
{\mathcal U}(t_0)={\mathcal U}(t_1)=\left(\begin{array}{cccc}0 & 0 & 0 & 1\\0 & 1 & 0 & 0\\0 & 0 & 1 & 0\\ 1 & 0 & 0 & 0
\end{array}\right), \ \ \ \ \ \ \ {\mathcal U}(t_2)={\mathcal U}(t_3)=\frac{1}{\sqrt2}\left(\begin{array}{cccc} 0 & 0 & 1 & -1\\0 & 0 & 1 & 1\\ 1 &1 & 0 & 0\\ -1 & 1 & 0 & 0
\end{array}\right)
\end{equation}
and ${\mathcal U}(t_n)=1\!\!1$ for any other $n$. Say the initial state is $\psi(t_0)=(|3\rangle+|4\rangle)/\sqrt2$.
Then easily
\begin{eqnarray}
\psi(t_{n\le 0})&=&            (|3\rangle+|4\rangle)/\sqrt2,
     \hspace{1cm}
\psi(t_1)=          (|3\rangle+|1\rangle)/\sqrt2,       \no
\psi(t_2)&=&           (|3\rangle+|4\rangle)/\sqrt2,
\hspace{1cm}
\psi(t_3)=           |2\rangle,       \no
\psi(t_{n\ge 4})&=& (|3\rangle+|4\rangle)/\sqrt2.
\end{eqnarray}
Hence
\begin{equation}
\psi_1(t_n)=\delta^1_n\  \frac{1}{\sqrt2},
\end{equation}
and Dolby's probability gives
\begin{equation}
\wp(t_n\  {\rm when}\ \   i=1) = \delta^1_n.
\end{equation}
But, accordingly with quantum mechanics, if we have detectors in
$i=1$ for all $t_n$, then at $t_1$ the state is projected with
probability 1/2 on $ |1\rangle$, and with probability 1/2 on $
|3\rangle$.  In both cases, the detector at $t_3$ will click with
probability 1/2, as a straightforward calculation shows.  Hence the
probabilities are: 1/4 that $t_1$ alone clicks, 1/4 that $t_3$ alone
clicks, 1/4 that $t_1$ and $t_3$ click, 1/4 that no detector click.
The ratio $N_{t\uparrow}/N_\uparrow$ is therefore 1/2, different
from Dolby's probability.  The point is of course that the very
presence of the detectors changes the wave function at later times,
and this is not taken into account by Dolby's expression.

Thus, neither the operational interpretation (i), nor the
operational interpretation (ii) work. What is then the meaning of
this probability?  We do not exclude that third consistent
operational definitions of Dolby's probability could be given, but
we do not yet see it. (On this problem, see \cite{Frank}.) Dolby
appears to be aware of this difficulty: in  \cite{Dolby:2004ak}, he
writes ``$\wp(a\  {\rm when}\ \  b)$ is best thought as representing
the proportion of the physical path on which $a$ and $b$ are
simultaneously true, divided by the proportion on which $b$ is
true". This is clearly intuitively okay, but in strident
contradiction with the fact that there is no way of measuring
``physical paths" in quantum mechanics.  The difficulty is that
Dolby's probability is defined as if the detectors did not affect
the quantum states; but this
contradicts the very physical contents of quantum theory.

For clarity, the difficulty we are raising is not that Dolby's
scheme cannot take into account situations with several
measurements. Dolby defines also probabilities, such as $\mathcal
P(a_1 \ {\rm when} \ b_1|a_2 \ {\rm when}\ b_2)$ that do so. The
point we are raising is that we do not understand how to interpret,
that is, how to concretely measure, Dolby's probabilities when the
$b$ variable in (\ref{dolby}) is not conventional time.

Thus, in spite of the attractiveness of the conditional probability
interpretation, we find that its foundation is not solid.  Therefore
we do not see, at present, how to take it as a general basis for
relativistic quantum mechanics.  The present work can be seen, to a
large extent, as an attempt to ameliorate the conditional
probability interpretation, but putting it on firmer grounds.

\section{Multi-event probability from the coupling of an apparatus}\label{apparatus}

\subsection{The basic idea}

It is perhaps the very central physical content of quantum theory
that certain questions cannot be combined by AND and OR, even if we
can do so in classical physics.  For instance, we can meaningfully
ask if an electron is in a certain region of space, and we can
meaningfully ask if an electron has a certain energy.   But asking
if an electron ``is in a region AND has a certain energy", makes no
sense in quantum theory,
beyond a certain precision. Position and energy are represented by
non-commuting operators in the quantum theory, and we can assign an electron
a definite position and a definite energy, but not
both. Bohr has emphasized that this is related to the fact that
electron manifests its position in a certain context of interaction
and it manifests its energy in a different context, but there is no
interaction context in which the two can be manifested together.
Position and energy are incompatible observables that cannot be measured
together, and therefore it is perfectly okay that we cannot
assign them joint probabilities.

At the light of this well known basic consideration, let us return
to the case of the measurement of the $s'$ and $s''$ events, in
the case {\bf\em (ii)} of Section \ref{3}, considered above.  For
concreteness, let us consider again the example in which $s'$
represent the spin being $\uparrow$ at a time $t'$ and $s''$
represents the spin being  $\leftarrow$ at a later time $t''$.
Since in the physical state space the two states $|\p s'\rangle$
and $|\p s''\rangle$ are not orthogonal, they cannot be
interpreted as (non-degenerate) eigenstates of commuting
operators.  We are led to the tentative idea that they
represent incompatible observables, to which we cannot assign
joint probabilities. After all, it is hard to say that two events
at different times can happen \emph{together}!

At first sight, this sounds as an elegant solution of the problem. But
at a moment of reflection, something is clearly missing:
Indeed, it \emph{does} make sense to ask the experimental question of
what is the probability for a spin system to have spin  $
\uparrow $ at time $t'$ AND spin  $\leftarrow
$ at a later time $t''$.  This is a statement that gets a precise
meaning in an appropriate measurement context.
In conventional QM, it can be dealt with by separating the two
measurements in time, and using the collapse algorithm to
compute joint probabilities.
Can we obtain the same probabilities without invoking the
collapse postulate?
As well known, the answer is yes, and follows from an analysis
of the experimental situation involved in the experiment in which
we measure the two spins at different times. The key is to bring the
apparatus that makes the measurement into the picture.

\subsection{Multi--event probability from the single--event one in nonrelativistic QM}

Let us see how this works in conventional QM.
Consider an initial state $| \psi \rangle$ at time $t$.
What is the probability of detecting the state
 $| \psi' \rangle$ at time $t'$ AND the state
 $| \psi'' \rangle$ at time $t''>t'$? This
 can be computed by projecting on  $| \psi' \rangle$ at time $t'$,
which gives
\begin{equation}
\wp_{\psi\Rightarrow \psi' \psi''}  = |\langle  \psi''  |U(t''-t')
\Pi_{\psi'} U(t'-t)|  \psi \rangle|^2.
\label{one}
\end{equation}
where $ \Pi_{\psi'} = | \psi' \rangle\langle \psi'  |$ and all
states are normalized $ \langle  \psi  |   \psi \rangle=1$. Now,
describe the apparatus measuring $\psi'$ as a two-state system which
is initially in a state $|No\rangle $, which interacts with the
system at the time $t'$, jumping to the state $|Yes\rangle $ if and
only if the state of the system is in the state  $\psi'$.   The
interacting dynamics is then given by the unitary evolution operator
$U_{\psi',t'}$ defined by
\begin{equation}
\langle  \psi'', A''  | U_{\psi',t'}(t''-t)|  \psi, A \rangle=
\langle  \psi''  | U(t''-t)|  \psi \rangle\ \delta_{A''A}
\end{equation}
for $t$ and $t''$ both larger or smaller than the interaction time
$t'$, and
\begin{equation}
\langle  \psi'', A''  | U_{\psi',t'}(t''-t)|  \psi, A \rangle=
\langle  \psi''  | U(t''-t')
\ \Big(\Pi_{\psi'} {\cal I}_{A''A}+(1-\Pi_{\psi'})\delta_{A''A}\Big)
U(t'-t)| \psi \rangle
\end{equation}
if $t''>t'>t$, where
\begin{equation}
{\cal I}_{AA'} \equiv \left(\begin{array}{cc} 0& 1\\
1 & 0\end{array} \right)
\label{defA}
\end{equation}
is a matrix that flips the apparatus state.    Then the question
we are interested in can be rephrased as follows: ``Given an
initial state $|\psi, No\rangle $  at time $t$, what is the
probability of measuring the state $|\psi'', Yes\rangle$ at time $t''$?"
Notice that ``the  system state is $|\psi''\rangle$" and ``the apparatus state is
$|Yes\rangle $", are \emph{compatible} statements in quantum theory: they
refer to orthogonal observables, both  at time $t''$, that are, and can be, measured
together!   In other words, the question can be captured by the
single-event probability amplitude
\begin{equation}
\wp_{\psi\Rightarrow \psi' \psi''}  = |\langle  \psi'',Yes  |U_{\psi',t'}(t''-t)|  \psi, No \rangle|^2.
\label{two}
\end{equation}

The same observation can also be illustrated as follows.  The probability of
a time--ordered sequence of events, such as (\ref{one}), can be written in the form
\begin{equation}
  \p_{\psi\Rightarrow \psi' \psi''}= \langle   \psi   |\Pi_{\psi', \psi''}  | \psi  \rangle
\end{equation}
where the history operator \cite{Hartle:1992as}
\begin{equation}
\Pi_{\psi', \psi''}= U(t-t')  \Pi_{\psi'} U(t'-t'') \Pi_{\psi''}
U(t''-t')  \Pi_{\psi'}  U(t'-t)
\end{equation}
is \emph{not} a projector.   But this same probability \emph{can} be
written as the expectation value of a projection operator if we enlarge the
state space to include the apparatus. In fact, (\ref{two}) gives
\begin{equation}
  \p_{\psi\Rightarrow \psi' \psi''}= \langle \psi, No   | \tilde\Pi_{\psi', \psi''}  | \psi, No  \rangle
\end{equation}
where the operator
\begin{equation}
\tilde\Pi_{\psi', \psi''}= U^\dagger_{\psi't'}\  \Pi_{\psi'', Yes}\
U_{\psi't'}
\label{paperino}
\end{equation}
\emph{is} a projector.  In fact, it projects on the state
$U^\dagger_{\psi't'}| \psi'', Yes\rangle$. That is, the probability
of a sequence of events can \emph{still} be obtained from the basic
probability postulate (\ref{singleeventP}).

\subsection{Multi--event probability from the single--event one in relativistic QM}

Let us translate the above observation in the language of general
relativistic quantum mechanics. An apparatus is now a two-state
system which is initially in a state $|No\rangle $, and interacts
with the system at the event $s'$, jumping to the state $|Yes\rangle
$ if and only if the event $s'$ happens. The question ``Given an
event $s$, what is the probability of detecting an event $s'$  AND
an event $s''$?" can be rephrased as follows: ``Given an event $|s,
No\rangle $, what is the probability of detecting the event  $|s'',
Yes \rangle$?" In other words, the question can be captured by a
\emph{single-event} probability amplitude
\begin{equation}
A_{s{\Rightarrow} (s' AND\  s'')}= A_{s, No{\Rightarrow} s'', Yes} =
\langle s'', Yes |\p_{s'}| s, No  \rangle, \label{twospinmeas}
\end{equation}
provided that the dynamical operator $\p_{s'}$ takes appropriately into account
the coupling between system and apparatus at the event $s'$.

It is clear that this strategy works in general, for arbitrary sequences of
measurements.  \emph{Sequences of incompatible measurements can
always be
reinterpreted as compatible measurements of apparatuses that have
interacted with the system.} In fact, it is clear that the only
operational meaning that we can ascribe to the probability that
something happen at time $t'$ AND something happen at a different
time $t''$ regards, in reality, the probability of a simultaneous
check of \emph{records} of (one at least  of) these two events.
Using this strategy, any probability for sequences of observations
can be reduced to a probability for eigenvalues of commuting
observables, and therefore be reduced to the single event
probability (\ref{singleevent}).

Where is the information about time-ordering gone?  It has been
coded into the specification of the interaction between the system
and the apparatus, namely in a new projection operator  $\p$ that includes the
interaction with the apparatus. The fact that the probability of a
sequence of events depends on the time ordering of the measurements
can be coded into the specification of the dynamics of the physical
interaction between the system and the apparatus. This observation
allows us to completely reduce multi-event probabilities to
single-event probability, and therefore to obtain all relevant
quantum mechanical probabilities from the single probability
postulate (\ref{singleevent}).

The reader might object that we have only shifted the difficulty
from the problem of defining the general formalism of quantum theory
to the problem of constructing a suitable dynamical operator
$\p_{s'}$ capable of capturing the abstract idea of a ``detection at
the event $s'$". We think this objection is ill-founded, for the
following reason.  In dealing with physics the problem is not to
describe abstract ideas, but to describe what we do concretely, and
to give predictions about concrete experimental situations. Each
concrete experimental situation has to be described by a specific
dynamics, and therefore by a specific $\p_{s'}$ operator. The
question is not whether or not a concrete system-apparatus
interaction describes or not an abstract idea of measurement: the
question is to find a formalism capable of describing and predicting
any concrete physical situation.  The problem of writing the correct
dynamics describing the situation at hand is a concrete problem in
the application of QM, not in the definition of its general
structure.  The general theory does not say what happens at
different times: for every physical situation it gives the
probability distribution for all the events, including those that we
may wish to view as records of previous events.

\subsection{A simple model}\label{model}

Let us illustrate this idea in the simple spin system model we have
used so far.  This shows how $\p$ is modified by the presence of the
apparatus. Recall that here ${\cal K}= \mathbb C^2\otimes L[\mathbb
R]$ spanned by a (generalized) basis of states $ | S, t \rangle,
S=\uparrow, \downarrow$.  Then ${\cal H}=\mathbb C^2$ and, choosing
the reference time $t_0=0$, $\p |S, t \rangle=
U^\dagger(t)|S\rangle$. We couple an apparatus with state space
$\mathbb C^2$, spanned by the states $ | A \rangle, A=Yes, No$.  The
kinematical Hilbert space becomes ${\cal K}= \mathbb
C^2\otimes\mathbb C^2\otimes L[\mathbb R]$ spanned by a
(generalized) basis of states $ | S, A, t \rangle$. The modified
dynamics, that includes an appropriate interaction is given, say, by
the operator $\p_{\uparrow t'}$ defined as
\begin{equation}
\langle   S'',
A'', t''  |\p_{\uparrow t'}|  S, A, t \rangle = U_{S''}{}^S(t''-t)\ \delta_{AA''}
\label{spinapparatus}
\end{equation}
if $t$ and $t''$ are both smaller or
larger than $t'$ (the interaction time); and
 \begin{equation}
 \langle   S'', A'',
t''  |\p_{\uparrow t'}|  S, A, t \rangle =
\sum_{S'=\uparrow,\downarrow}U_{S''}{}^{S'}(t''-t')\ \;
\Big(\delta_{S'}^\uparrow\ {\cal I}_{A''A}+ \delta_{S'}^\downarrow\
{\delta}_{A''A}\Big)\; U_{S'}{}^S(t'-t)
\label{spinapparatus2}
\end{equation}
 if $t<t'<t''$. Notice that if and only if the spin is up, the
apparatus flips state at time $t'$. That is, the dynamics
defined by $\p_{\uparrow t'}$ is the one that describes precisely an
apparatus measuring $\uparrow$ at time $t'$.

Equivalently, we can write (\ref{spinapparatus2}) posing (always for
$t<t'<t''$)
 \begin{equation}
  \p_{\uparrow t'}
 = \p\otimes \mathbb{I}_{app} \ + \
 \p\  \Pi_{\uparrow t'} \p  \otimes   \tilde{\cal I}
\end{equation}
where $\langle A |\mathbb{I}_{app}|A'\rangle=\delta_{AA'}$ and
$\langle A |\tilde{\cal I}|A'\rangle=\tilde{\cal I}_{AA'}$ is
\begin{equation}
\tilde{\cal I}_{AA'} \equiv \left(\begin{array}{rr} -1& 1\\
1 & -1\end{array} \right)
\label{defA2},
\end{equation}
which emphasize the fact that $\p_{\uparrow t'}$ is equal to
$\p$ plus an interaction term at $s'$.

We can now ask what is the probability of detecting, say, the state
$|\tilde s'\rangle=|\!\!\leftarrow ,Yes, t'' \rangle$ in the
$\mathcal K$ space of the coupled (system+apparatus) system, given
that the state $|\tilde s\rangle=|{\rightarrow} ,No, t \rangle$ was
detected. Here $\tilde s'$ and $\tilde s$ are events of the coupled
(system+apparatus) system. Then (\ref{singleevent}) and
(\ref{spinapparatus2})  give the correct result obtained from
conventional QM.  For instance, if, say, $U(t)=1\!\! 1$ (cf.\
(\ref{sequence})),
 we obtain
 $\wp_{s{\Rightarrow}(s'\, {\scriptscriptstyle
\rm AND}\, s'')} =1/4$ if $t<t'<t''$ and $\wp_{s{\Rightarrow}(s'\,
{\scriptscriptstyle \rm AND}\, s'')}=0$ if $t<t''<t'$, to be
compared with \eqref{jenny}.

An example of application of this idea to a system genuinely without unitary
time evolution will be given in the companion paper \cite{model}.

\section{The meaning of probability}\label{probabilita}

Before concluding, we discuss here a common objection to the
assignment of probabilities to individual events, without reference
to equal time surfaces.

A probability is meaningful only if it assigned to an event $s$ out
of a set of alternatives ${\cal S}=\{s_1, s_2, ... \}$. In
non-relativistic QM, the probability of finding a particle in a
certain spatial region $R$ is usually understood as the probability
of finding the particle in this region out of the alternatives given
by the possibility of finding the particle {\em in other spatial
regions at the same time}.   That is, if  $s=(R,t)$, then  ${\cal
S}={\cal S}_t=\{(R_n,t)\}$ where the $R_n$ are an ensemble of
regions that cover an equal time surface.  In a general relativistic
context, there is nothing that singles out the equal time surface.
Does this imply that the probability $\mathcal P_s$ has no meaning
in a general relativistic context?

The answer is no, for the following reason.   There exists an
alternative interpretation of the probability $\wp_s$ that does not
require the set of alternatives ${\cal S}_t=\{(R_n,t)\}$.
Consider a detector which is
active in the space region $R$ at time $t$. This detector has a
finite probability of detecting the particle {\em and a finite
probability of not detecting the particle}.  These two alternatives
define a simpler set \begin{equation} {\cal S}_{dual}=\{{\rm\emph{Detection, Non
detection}} \}. \label{alternatives} \end{equation} We can then interpret
$\wp_s$ as the probability that detector detects the particle, out
of the two alternatives (\ref{alternatives}).  This does not require
the equal time surface ${\cal S}_t$ to play any role.

In fact, a moment of reflection will convince the reader that that
this is what we truly mean by $\wp_s$ in any realistic quantum
mechanical measurement.   If the set of alternatives  ${\cal S}_t$
was the relevant one, any position measurement would only be
consistent if, at the same time $t$, there were detectors all over the universe, all measuring
whether the particle is there!  This is
\emph{not} what we do when we measure if the particle is in a
certain region at a certain time. What we do is to have a detector
only in the region of concern and \emph{interpret} the case of non
detection as implying that the particle would have been detected by
one detector elsewhere.  We can do so, because in ordinary
non-relativistic quantum mechanics, we have the remarkable property
that
\begin{equation}
           \sum_n\ \ \wp_{(R_n,t)} = 1. \label{consistency}
\end{equation} But this is a specific property of the dynamics of a particle,
not a condition for $\wp_{R_n,t}$ to be defined.  $\wp_{R_n,t}$ is
defined by itself, and the probability normalization condition is
simply
\begin{equation}
           \wp_{(R,t)}+\wp_{\rm\scriptsize\emph{Non detection}}=
           \wp_{\rm\scriptsize\emph{Detection}}+\wp_{\rm\scriptsize\emph{Non detection}} = 1.
\end{equation}

Let us express the same idea in more mathematical terms. Let $|x\rangle_t$
be an eigenvectors of the Heisenberg position operator $X(t)$, that is
$X(t)|x\rangle_t= x|x\rangle_t$.  The state
 $|x\rangle_t$ is, of course, \emph{also} an eigenvector of the
proposition operator
\begin{equation}
P_{xt}=|  x\rangle_t\; {}_t\langle x| .
\end{equation}
 The two operators
are related by the spectral decomposition
\begin{equation}
X(t)=\int dx \ x\
  P_{xt} \label{spectral}
  \end{equation}
 in nonrelativistic QM, but they are independently defined and
they have each a physical interpretation. The operator $X(t)$
describes an ensemble of detectors covering the entire space
and measuring where is the particle at time $t$. Its outcome
is a real number $x$.  The operator $P_{xt}$ describes a single
detector that detects whether or not the particle is in $x$ at time
$t$.  Its outcome is a single bit: either $YES$ or $NO$.
Now the relation (\ref{spectral})
implies (\ref{consistency}): if this relation is not available, the
proposition operator $  P_{xt}$ \emph{is still well defined}, and still
defines consistent probabilities for this detector outcomes.

\section{Perspectives}\label{fine}

The main result of this work is to show that the single probability
postulate (\ref{singleevent}) of general relativistic quantum
mechanics is capable of giving all probabilities of conventional
quantum mechanics, including the probabilities for \emph{sequences}
of events, which are usually computed by means of the projection
algorithm. This is achieved by exploiting the freedom of moving the
quantum/classical boundary, emphasized by von Neumann, and assuming
that (in the \emph{non-relativistic case}) the evolution of the
system+apparatus is always unitary.

A number of points deserve to be better understood. In particular:
\emph{(i)} The use of this formalism in less trivial systems, where
the complications connected to continuous spectrum operators and
infinite volume gauge groups are more severe. \emph{(ii)} The
possibility of associating probabilities to arbitrary continuous
regions of $\cal C$ \cite{Marolf:1993yx,Marolf:2002ve}. This is
related to the well known ``time of arrival" problem (see for
instance \cite{arrival} and references therein). \emph{(iii)} The
extension of these ideas to field theory, and in particular the
connection between this formalism and the boundary formalism
\cite{book,boundary} which is presently used  \cite{scattering} to
compute probability amplitudes in background independent quantum
gravity, in the context of loop quantum gravity
\cite{book,lqg,alex}. \emph{(iv)} Eventually, we would like to apply
this formalism to physical situations where the the assumption of
the existence of a background geometry breaks down (for instance,
see \cite{Rovelli:1996ti}).   We leave these issues for further
developments.

A number of tentative considerations following from the present
result can nevertheless be attempted.

\emph{(i)} \emph{Time ordering} does not appear to be a fundamental
structure required for the definition of quantum theory and the
calculation of its probability amplitudes.  In our opinion, this
reinforces the hypothesis that the fundamental theory of nature can
be formulated in a timeless language \cite{Rovelli:1990ph}, and that
temporal phenomena could be emergent \cite{time}.

\emph{(ii)} In the generally covariant context, dynamics can
be entirely expressed in terms of \emph{Dirac observables}.
Indeed, notice that the probability of a sequence of measurements
can be written as in equation (\ref{singleeventnorm}), namely
as the expectation value of the projection operator $\Pi_s$
defined in (\ref{singleeventP}), or (\ref{paperino}).   This
operator  \emph{is} a Dirac observable of the extended
system that includes the measuring devices.

In the present context, this is the answer to the long-standing problem of the description
of dynamics in the ``frozen-time''  formalism; namely in the
Dirac's quantization of a system whose dynamics is expressed by
constraints \cite{Kuchar:1991qf,Isham:1992ms}.  Dynamics is
coded into (non-commuting) Dirac observables defined in terms
of sets of interactions between (what we call) the system
and (what we call) the measuring devices.

\emph{(iii)} The discussion above bears also, indirectly, on the
discussion on the interpretation of quantum mechanics, and on
the nature of the quantum collapse.

In some interpretations of quantum mechanics, the wave function is
considered a real entity that evolves unitarily, except at
measurement time, when it undergoes a sudden change.  In particular,
some interpretations make the hypothesis that this  ``collapse" is a
real physical phenomenon whose peculiar nonlocal dynamics is not yet
understood. If this is the case, the full freedom of moving the
quantum/classical boundary is broken, because once the collapse has
happened no more interference between the two ``branches" of a
measurement outcome is possible, even in principle.  If this is the
case, the strategy adopted here is not viable in general, because it
assumes, instead, that no true physical collapse happens at anytime.

In some others interpretations, the wave function, or the ``quantum
state", is not considered as a real entity. Rather, only quantum
``events" are considered real, and probabilities like $|\langle
s'|s\rangle|^2$ are directly interpreted as conditional
probabilities for these events to happen. In particular,  in
\cite{relational} these quantum events are assumed to happen at
interactions between systems, and to be real only with respect to
the interacting systems themselves. From this perspective, there is
no specific physical phenomenon corresponding to a quantum collapse,
and the strategy considered here is viable. With respect to an
external system, what happens at the interaction between system and
apparatus is not a sudden change in a hypothetical real  ``state",
but simply an entanglement between the probabilities of various
outcomes of observations on the system or the apparatus.  We refer
to \cite{relational}, and Section 5.6 of \cite{book} for a
discussion of this point of view.

\emph{(iv)} To our knowledge, the only complete general covariant
formalism for quantum theory alternative to the one we have
discussed here is Hartle's generalized quantum mechanics
 \cite{Hartle:1992as}.   We find an interesting convergence between
Hartle's covariant sum over histories and our results. Within
Hartle's generalized quantum mechanics, probabilities for sequences
of events can be expressed by means of history operators.  What we
have argued here is that the mean value of a history operator can be
re-expressed as the mean value of a conventional projection operator
on the joint system+apparatus Hilbert space.  This relations will be
better discussed and illustrated in the companion paper
\cite{model}.

\emph{(v)} Finally, in our opinion the result presented here reinforce the idea
that quantum mechanics admits a very simple and straightforward
generalization which is fully consistent with general relativity.
And therefore that the contradictions between quantum
theory and general relativity might be only apparent.

\end{document}